\documentstyle[twocolumn,prl,aps]{revtex}

\begin{document}
\twocolumn[
\hsize\textwidth\columnwidth\hsize\csname @twocolumnfalse\endcsname

\title{ 
Energy landscape, two-level systems and entropy barriers in
Lennard-Jones clusters
	}
\author{ 
	G.Daldoss, O.Pilla and G.Viliani
	}
\address{ 
	  Dipartimento di Fisica and Istituto Nazionale di Fisica della 
	  Materia, Universit\`a di Trento, I-38050 Povo, Trento, Italy
	  }
\author{G.Ruocco}
\address{ 
	Dipartimento di Fisica and Istituto Nazionale di Fisica della
        Materia, Universit\`a dell'Aquila, I-67100 Coppito, L'Aquila, Italy
        }
\maketitle

\begin{abstract}
We develop an efficient numerical algorithm for the identification of a
large number of saddle points of the potential energy function of Lennard-
Jones clusters. Knowledge of the saddle points allows us to find many
thousand adjacent minima of clusters containing up to 80 argon atoms and to
locate many pairs of minima with the right characteristics to form two-level
systems (TLS). The true TLS are singled out by calculating the ground-state
tunneling splitting. The entropic contribution to all barriers is evaluated
and discussed. 
\end{abstract}
]
The use of an effective potential energy to describe the dynamical
properties of disordered systems has become a very powerful tool of
investigation. In particluar, there is
broad consensus on the idea that important relaxational processes can be
described by transitions between adjacent potential-energy minima in the
multidimensional configuration space. The latter, however, is so 
complex even for systems containing a moderate number of atoms, that
singling out relevant pairs of minima has required important numerical
efforts and skill by many authors \cite{sw,ws,s,hs1,hs2,hs3,heu}. One of the
goals of this kind of research is the identification of pairs of minima
suitable to produce two-level systems (TLS); these are the pairs of nearly
degenerate minima separated by a barrier that, following tunneling, produces
a total splitting of the order of $\approx <$ 1 K, thus explaining the
anomalous low-temperature specific heat and thermal conductivity of glasses
\cite{ahv,ph,zp}. 
\par
Identifying adjacent minima is not simple by molecular dynamics methods; it
is clear that the key stationary points are not the minima but rather the
(first-order) {\it saddle points}, descending from which adjacent minima are
automatically found. But there are other reasons why the search for saddle
points looks desirable, which are in a way or another related to the fact
that the minimum-minimum relaxation process takes place along a classical
path that passes close to the saddle, i.e. the least-action path. 
\par
The first reason is that evaluating the TLS ground state splitting is a
quantum mechanical problem, but the solution of the Schroedinger equation in
3$N$ dimensions is out of question and one is forced to think of ways of
drastically reducing the dimension of the problem, possibly to one. It is
therefore necessary to have an indication about how much and under which
circumstances this reduction is feasible; as we will discuss later this kind
of information requires a detailed knowledge of the path along which
tunneling is considered to take place and of the potential energy along it.
\par
The second reason, which is related to the first one, is that relaxation
processes of all kinds, either the low-temperature ones that produce TLS or
the high-temperature ones that govern the glass transition, in principle are
controlled not only by energy barriers, but also by {\it entropy} barriers,
i.e. by the topological difficulty of proceeding from one minimum to another
through a narrow route even in the absence of potential energy barriers
\cite{mot,ritort,ritfranz}. Note that entropy barriers would also affect
the quantum mechanical
splitting of the ground state in much the same way as energy barriers do
\cite{franci1}. 
\par
Heuer \cite{heu} has recently studied the potential-energy topology of
systems containing 32 atoms placed in a box with periodic boundary
conditions and interacting through a Lennard-Jones-like potential introduced
in ref. \cite{sw}.
At a numerical density $\rho=1$ in units of the nearest neighbor distance,
the author finds 367 minima of different energy, 
which become 75 by decreasing the box size by $\approx$2\%. These values are
exceedingly small when compared to the $\approx 10^{11}$ different minima
expected for a Lennard-Jones cluster of 32 atoms \cite{dpv}: the boundary
conditions have a dramatic effect on the number of minima (and probably of
all kinds of stationary points). Heuer \cite{heu} notes that performing the
simulation at variable box length produces many more minima; in view of this
situation, we think it is very important to search for the stationary points
in the absence of any boundary condition, i.e. for free clusters. The reason
is that even periodic boundary conditions with variable density are likely
to introduce spurious correlations among different boxes,
 that might affect the results.
The price to be paid is the presence of surface effects which (like the
correlations of the boundary conditions case) become less and less important
as the system size is increased. 
\par 
The systems we studied are Ar clusters with up to $N=$80 atoms, interacting
through the Lennard-Jones potential
$$
        V({\bf r})=4\epsilon \sum_i \sum_{j>i}[(\frac{\sigma}{r_{ij}})^{12} -
(\frac{\sigma}{r_{ij}})^{6}]
$$
with $\sigma=3.405$ $\AA$ and $\epsilon/k_B$=125.2 K. 
\par
Since we think of
our clusters as parts of a glass, in which case translations and
rotations are forbidden, we decided to set six coordinates to zero (one atom
is fixed in the origin, one on the $x$ axis and one on the $xy$ plane) in
order to eliminate translations and rotations. 
Free translations of the system as a whole are decoupled by other degrees of
freedom, but in non-rigid bodies
like clusters rotations and vibrations are coupled. Therefore, the above
conditions leave the potential energy of the stationary configurations
 unaffected because it depends only on mutual distances, but
the elimination of rotations affects vibrational frequencies by
altering the effective mass. We do not expect any systematic bias of the
results because the coordinates to be fixed are chosen at random.
\par
The saddle points were found with the procedure described in detail in ref.
\cite{dpv}, and which consists of the following steps: ($i$) descent towards
a minimum by the conjugate gradient method, starting from a randomly chosen
configuration; ($ii$) ascent towards the vicinity of a saddle following the
eigenvector corresponding to the minimum eigenvalue of the Hessian; this
requires diagonalization of the Hessian at each step; ($iii$) once the
potential energy along the path of the previous item starts decreasing, we
take the corresponding configuration as the starting point for a Newton-
Raphson stationary point search \cite{numrec}. This means that the non
linear system of equations
\begin{equation}
	\frac{\partial V({\bf r})}{\partial x_i} = 0
\end{equation}
is solved by successive iterations for $i$=1...($3N$-6). The steps ($i-ii$)
are required in order that the Newton procedure converges; the iteration
ends up in a first order saddle in about 30\% of cases and the saddle is
located with extreme accuracy. The next step is a minimum-eigenvalue
descent on both sides of the saddle, which provides approximate {\it
adjacent} minima that are subsequently fed into the Newton algorithm for
accurate location. With this numerical procedure we found many thousand
minimum-saddle-minimum triplets for several values of $N$ in the range 6-80.
The number of triplets found and of those with asymmetry $\Delta$<1 K, are
reported in the second and third columns of Table I, respectively.
The last 2 columns of the Table entail the
evaluation of the ground state splitting, which will be discussed later.
\par
The first check we performed was to look for correlations between the
asymmetry and the barrier height; contrary to previously reported results on
periodic systems \cite{hs1,franci2} we found no correlation. The pairs of
minima
were further characterized by their euclidean distance, $d$, and by the
participation number, $N_p=d^2/d_{max}^2$, where $d_{max}$ is the
displacement of the atom that moves most. The distribution of $d$ is peaked
at $\approx 4\sigma$ for $N=13-42$ and at $\approx 6\sigma$ for $N=80$, with
tails extending up to $\approx 15 \sigma$. These tails reflect the free-
surface effects that we know affect the clusters, but interestingly enough
the TLS pairs have in all cases $d < \approx 2\sigma$. This is convenient
for an a priori selection of candidate TLS. The participation number for
$N=42,80$ is reported in Fig. 1. In column ($b$) of Table I we report the
number of candidate TLS, i.e. those triplets that have both $\Delta$<1 K and
the barrier higher than the lowest vibrational eigenvalue of either minimum;
the distribution of their participation number is not very different from
that of all pairs for
$N \le 42$, while for $N=80$ we have too few pairs to draw any definite
conclusion. 
\par
In order to identify the TLS, we need the total splitting of the ground
level, resulting from asymmetry, $\Delta$, and tunneling, $\delta_T$.  
As mentioned above, the evaluation of the latter is a quantum
mechanical issue and can be solved relatively easily only if the many-
dimensional dynamics can be reduced to a 1-dimensional one. If this is at
least approximately feasible, then the computationally most convenient way
of evaluating $\delta_T$ is to use the semiclassical WKB approximation
\cite{frfr,landau} that provides an explicit expression in terms of the
potential energy profile $V(x)$:
\begin{equation}
	\delta_T=\frac{\hbar \omega}{\pi} D^{1/2}
\end{equation}
with $D=[1+ {\text exp}(2S)]^{-1}$ and the action integral
$S=\frac{1}{\hbar} \int^b_{a} \{ m[2V(x) - \hbar \omega] \} ^{1/2} dx $ is
evaluated between the classical turning points $a$ and $b$; $\omega$ is the
classical frequency in one well. The total splitting is then obtained as
$\delta \approx \sqrt{\Delta^2 + \delta_T^2}$. As mentioned, in principle it
is possible to use equation (2) for evaluating the tunneling splitting only
if the Schroedinger equation can be separated into independent ones for
each degree of freedom \cite{schiff}; we shall assume that (2) can be
employed under the less restrictive condition that the {\it relevant}
classical degree of freedom, i.e. the coordinate along the least action
path, is decoupled by the
remaining ones. Demichelis et al \cite{franci1,franci2} first found that
this actually occurs for argon with periodic boundary conditions, and the
same has been verified for candidate TLS in clusters \cite{giorgia}: the
higher vibrational eigenvalues, obtained by diagonalizing the dynamical
matrix at configurations along the least-action path, do not change
appreciably along the path itself, while the ground one obviously does.
Under these circumstances the use of Eq. (2) is reasonably justified
\cite{schiff}. The least-action path was determined by minimizing the action
integral $S$ following the procedure of ref. \cite{franci1}, and by taking
as initial path the one starting from the saddle and reaching the minima by
a minimum eigenvalue search. The action integral was evaluated also along
the straight line connecting the minima; for the TLS the straight-line
integral turns out to be only a factor 2-5 larger than the least-action one.
\par
In the last column of Table I (TLS) we report the number of triplets that
have a total (asymmetry plus tunneling)  splitting in the ground state
$\delta <$ 1 K,
and
thus have all the characteristics of TLS \cite{nota1}. We see that
interestingly enough the
category of TLS
almost coincides with that of column ($b$). But most remarkably, it seems
that $n$(TLS)/$n$(Total) is roughly of the order of $10^{-3}$ and apparently
tends to increase with $N$; if this trend should be confirmed it would mean
that surface effects {\it do not} favor TLS, so that the failure to observe
them in periodic systems of comparable \cite{heu} or larger \cite{franci2}
size should probably be ascribed either to correlations or to density. In
any case, visual inspection of the TLS relative to $N=80$ shows that surface
effects are still very important because in all 3 cases there is at least
one surface atom that moves most. For $N=42,80$ we verified that TLS are
isolated, in the sense that we did not find any low-barrier route that
connects one minimum of the pair to a third almost degenerate minimum. This
was to be expected since only $\approx 1/100$-th of the pairs are nearly
degenerate (Table I, column ($a$)).
\par
The entropic contribution to classical relaxation was considered along the
following lines. The classical one-dimensional transition rate between
minima separated by a barrier $E_b$ is
\begin{equation}
        k=\frac{\omega}{2 \pi} {\text exp}(-\frac{E_b}{K_BT})
\end{equation}
where $\omega$ is the vibration frequency in the potential well; in many
dimensions and under some assumptions concerning thermodynamic equilibrium
and the absence of back crossing, it is found that the entropic-barrier
effect of degrees of freedom other than the single considered path can be
accounted for by the following substitution in the pre-exponential factor of
(3) \cite{rice,glyde,hanggi}:
\begin{equation}
        \omega \rightarrow \Pi_i \omega_i^M /\Pi'_j \omega_j^S 
        =\frac{\omega}{R}
\end{equation}
Here $\omega_i^M$ and $\omega_j^S$ are the vibrational eigenfrequencies at
the minimum and at the saddle respectively, and the prime indicates that the
imaginary frequency at the saddle point has to be omitted from the product.
\par
The entropic ratio $R$ has been calculated for all
minimum-saddle-minimum triplets for $N=$ 42 and 80, and is reported
for $N=$ 80 in fig. 2 as a function of the barrier.
For $N=$42 the distribution is qualitatively
similar but much more scattered. As can be seen, the  majority
of the triplets have $R$ in the range 1-$10^{-2}$, with a high-barrier
tail extending down to $R \approx 10^{-6}$. Only relatively few triplets
(0.45\%) have $R>1$ (but in any case not exceeding the value 10); for
$N=$42 this fraction becomes 0.75\%. The present
results seem to indicate that entropic barriers are not
especially active 
in slowing down the relaxation dynamics, and that the effect
may become less important with increasing $N$. This result is contrary
to what found recently for a different model of fragile glass
with periodic boundary conditions \cite{dasgu}: the role of
boundary conditions on the results of simulations on relatively
small systems definitely deserves further investigation.
\par
For the
TLS, in agreement with the reported independence of the eigenvalues
other than the ground one on the position along the least-action
path \cite{franci2}, we find $R \approx$ 1.
Considering that for $N=$80 $R$ is the ratio of two products of 234
numbers each, the result $R \approx 1$ is remarkable and shows that TLS
belong to a rather peculiar class of triplets.
In this case, entropic barriers do not
play a relevant role in determining the quantum splitting of the ground
level, though in general they quench the splitting like ordinary potential
barriers.
\par
In summary, TLS have been shown to exist in Lennard-Jones clusters and
have been characterized; the order of magnitude of the ground level quantum 
splitting has been evaluated by means of the semiclassical 1-dimensional WKB
formula; topological hindrances do not play a significant
role in determining this splitting. Moreover, the relaxation dynamics
of the vast majority of minimum-saddle-minimum triplets appears to be
little
affected by entropic barriers. As far as we are aware, this kind of
information on tunneling and relaxation processes has not been provided
before, and it was made possible by the use of a numerical procedure that
produces a large amount of adjacent minima. The quantitative application of
these results to glasses requires great caution, because surface effects are
certainly present in our clusters. However, the influence of boundary
conditions and density on the properties of the landscape \cite{heu} is
also very great and the present zero-pressure and correlation-free results
may certainly help in getting a closer insight into the problem.
\par
We are grateful to A. Ranfagni and D. Mugnai for very useful discussions.

\vskip 1cm
\noindent
\narrowtext
\begin{table*}
\caption
{Number of pairs of adjacent minima found for $N=6 - 80$. $\Delta$ is the
asymmetry and $\delta$ the total splitting of the ground state
(see text). (a): $\Delta<1$ K; (b): Same as (a) but with barrier
higher than minimum vibrational eigenvalue; (TLS): same as (b) but with
total splitting $\delta<1$ K and tunneling splitting $\delta_T>10^{-15}$ K
(see ref. [20]). }
\begin{tabular}{ccccc} 
N& Total &(a)&(b)&TLS\\
 \hline
6&6&0&0&0\\
8&61&0&0&0\\
9&181&4&0&0\\
10&414&2&0&0\\
13&3416&41&3&3\\
15&4652&43&6&4\\
18&11412&150&19&15\\
29&9176&63&18&13\\
42&2878&31&11&7\\
80&1290&9&3&3\\
\end{tabular}
\end{table*}

FIGURE CAPTIONS\\
\vskip 1cm
Fig. 1. Participation number for clusters with $N=$ 42 (squares) and 80
(circles).
\vskip 1cm
Fig. 2. Entropic ratio $R$, as defined in Eq. (4), for $N=$ 80 as a
function of the saddle-minimum energy barrier; full circles: upper minimum;
open squares: lower minimum.


\begin{references}
\bibitem{nota1}
There exist also pairs of minima with very little or no asymmetry but
characterized by huge action integrals leading to tunneling probabilities of
the order of less than 1 event per day; these involve large displacements of
surface atoms and have been discarded.
\end{references}
\end{document}